\documentclass[preprint,authoryear,12pt,5p,times]{elsarticle}
\pdfoutput=1

\usepackage{graphics}
\usepackage{graphicx}

\usepackage{amssymb}
\usepackage{mathtools}
\usepackage{subfigure}
\usepackage{color}
\usepackage{listings}
\usepackage{parallel}
\usepackage{url}

\newsavebox{\mylistingbox}

\definecolor{light-gray}{gray}{0.95}
\newcommand{\arcsec}{\mbox{$^{\prime\prime}$}}

\journal{Astronomy and Computing}

\begin{document}

\begin{frontmatter}


\author[cfa]{Navtej Singh\corref{cor1}\fnref{fn1}}
\ead{n.saini1@nuigalway.ie}

\author[cfa]{Lisa-Marie Browne}
\ead{l.browne1@nuigalway.ie}

\author[cfa]{Ray Butler}
\ead{ray.butler@nuigalway.ie}

\address[cfa]{Centre for Astronomy, School of Physics, National University of Ireland - Galway, University Road, Galway, Ireland}

\cortext[cor1]{Corresponding author}
\fntext[fn1]{Telephone: +353-91-492532, Fax: +353-91-494584}


\title{Parallel Astronomical Data Processing with Python: \\ Recipes for multicore machines}




\begin{abstract}
High performance computing has been used in various fields of astrophysical research. But most of it is implemented on massively parallel systems (supercomputers) or graphical processing unit clusters. With the advent of multicore processors in the last decade, many serial software codes have been re-implemented in parallel mode to utilize the full potential of these processors. In this paper, we propose parallel processing recipes for multicore machines for astronomical data processing. The target audience are astronomers who are using Python as their preferred scripting language and who may be using PyRAF/IRAF for data processing. Three problems of varied complexity were benchmarked on three different types of multicore processors to demonstrate the benefits, in terms of execution time, of parallelizing data processing tasks. The native multiprocessing module available in Python makes it a relatively trivial task to implement the parallel code. We have also compared the three multiprocessing approaches - Pool/Map, Process/Queue and Parallel Python. Our test codes are freely available and can be downloaded from our website. 
\end{abstract}


\end{frontmatter}


\section{Introduction}
\label{intro}
In 1965, Gordon Moore predicted that the number of transistors in integrated circuits would double every two years \citep{moore_cramming_1965}. This prediction has proved true until now, although semiconductor experts\footnote{From the 2011 executive summary of International Technology Roadmap for Semiconductor (http://www.itrs.net/links/2011itrs/2011Chapters/2011ExecSum.pdf)} expect it to slow down by the end of 2013 (doubling every 3 years instead of 2). The initial emphasis was on producing single core processors with higher processing power. But with increasing heat dissipation problems and higher power consumption, the focus in the last decade has shifted to multicore processors - where each core acts as a separate processor. Each core may have lower processing power compared to a high end single core processor, but it provides better performance by allowing multiple threads to run simultaneously, known as thread-level parallelism (TLP). At present, dual and quad core processors are common place in desktop and laptop machines and even in the current generation of high end smart phones. With both Intel \citep{garver_new_2009} and AMD\footnote{Advanced Micro Devices} working on next generation multicore processors, the potential for utilizing processing power in desktop machines is massive. However, traditional software for scientific applications (e.g. image processing) is written for single-core Central Processing Units (CPU) and is not harnessing the full computational potential of the multicore machines. 

Traditionally, high performance computing (HPC) is done on supercomputers with a multitude of processors (and large memory). Computer clusters using commercial off the shelf (COTS) hardware and open source software are also being utilized \citep{szalay_extreme_2011}. And recently graphical processing unit (GPU) based clusters have been put to use for general purpose computing \citep{strzodka_scientific_2005,belleman_high_2008}. The advent of multicore processors provides a unique opportunity to move parallel computing to desktops and laptops, at least for simple tasks. In addition to hardware, one also needs unique software protocols and tools for parallel processing. The two most popular parallel processing protocols are Message Passing Interface (MPI) and OpenMP. MPI is used on machines with distributed memory (for example - clusters) whereas OpenMP is geared towards shared memory systems.

Parallel computing has been used in different sub-fields of astrophysical research. Physical modeling and computationally intensive simulation code have been ported to supercomputers. Examples include N-Body simulation of massive star and galaxy clusters \citep{makino_grape-4:_1997}, radiative transfer \citep{robitaille_hyperion:_2011}, plasma simulation around pulsars, galaxy formation and mergers, cosmology etc. But most of the astronomical image processing and general time consuming data processing and analysis tasks are still run in serial mode. One of the reasons for this is the intrinsic and perceived complexity connected with writing and executing parallel code. Another reason may be that day to day astronomical data processing tasks do not take an extremely long time to execute. Irrespective of this, one can find a few parallel modules developed for astronomical image processing. The cosmic ray removal module CRBLASTER \citep{mighell_crblaster:_2010} is written in C and based on the MPI protocol, and can be executed on supercomputers or cluster computers (as well as on single multicore machines). For co-addition of images, \citet{wiley_astronomy_2011} proposed software based on the MapReduce\footnote{Model to process large data sets on a distributed cluster of computers} algorithm, which is geared towards processing terabytes of data (for example - data generated by big sky surveys like the SDSS\footnote{SDSS: Sloan Digital Sky Survey [http://www.sdss.org/]}) using massively parallel systems.

In this paper, we have explored the other end of the spectrum - single multicore machines. We are proposing a few recipes for utilizing multicore machines for parallel computation, to perform faster execution of astronomical tasks. Our work is targeted at astronomers who are using Python as their preferred scripting language and may be using PyRAF\footnote{PyRAF is a product of the Space Telescope Science Institute, which is operated by AURA for NASA.} or IRAF\footnote{IRAF is distributed by the National Optical Astronomy Observatories, which are operated by the Association of Universities for Research in Astronomy, Inc., under cooperative agreement with the National Science Foundation.} for image/data processing and analysis. The idea is to make the transition from serial to parallel processing as simple as possible for astronomers who do not have experience in high performance computing. Simple IRAF tasks can be rewritten in Python to use parallel processing, but rewriting the more lengthy tasks may not be straightforward. Therefore, instead of rewriting the existing optimized serial tasks, we can use the Python multiprocessing modules to parallelize iterative processes. 

In Section \ref{pdp}, we introduce the concept of parallel data processing and the various options available. Python multiprocessing in discussed in Section \ref{python_multiprocessing} with emphasis on native parallel processing implementation. Three different astronomical data processing examples are benchmarked in Section \ref{examples}. In Section \ref{discuss}, we discuss load balancing, scalability, and portability of the parallel Python code. Final conclusions are drawn in Section \ref{conclude}.


\section{Parallel Data Processing}
\label{pdp}
Processors execute instructions sequentially and therefore, from the initial days of computers to the present, most of the applications have been written as serial code. Generally coding and debugging of serial code is much simpler than parallel code. However, debugging is an issue only for parallel programs where many processes depend on results from other processes -  whereas it is not an issue while processing large datasets in parallel. Moving to parallel coding not only requires new hardware and software tools, but also a new way of tackling the problem in hand. To run a program in parallel, one needs multiple processors/cores or computing nodes\footnote{The terms processors and computing nodes will be used interchangeably in the rest of the paper.}. The first question one asks is how to divide the problem so as to run each sub-task in parallel.

Generally speaking, parallelization can be achieved using either \textit{task parallelization} or \textit{data parallelization}. In task parallelism, each computing node runs the same or different code in parallel. Whereas, in data parallelism, the input data is divided across the computing nodes and the same code processes the data elements in parallel. Data parallelism is simpler to implement, as well as being the more appropriate approach in most astronomical data processing applications, and this paper deals only with it.

Considering a system with N processors or computing nodes, the speedup that can be achieved (compared to 1 processor) can be given as:

			\begin{equation} \label{eq:speedup}
				S = \frac{T_1}{T_N}, 
			\end{equation}

where $T_1$ and $T_N$ are the code runtime for one and N processors respectively. $T_N$ depends not only on the number of computing nodes but also on the fraction of code that is serial. The total runtime of the parallel code using N processors can be expressed using Amdahl's law \citep{amdahl_validity_1967}:

			\begin{equation} \label{eq:runtime}
				 T_N = T_S + \frac{T_P}{N} + T_{sync} 
			\end{equation}

where $T_S$ is the execution time of the serial fraction of the code, $T_P$ is the runtime of code that can be parallelized, and $T_{sync}$ is the time for synchronization (I/O operations etc.). The efficiency of the parallel code execution depends a lot on how optimized the code is, i.e. the lower the fraction of serial code, the better. If we ignore synchronization time, theoretically unlimited speedup can be achieved as $N\to\infty$ by converting the serial code to completely parallel code. More realistically, $T_{sync}$ can be modelled as $K * ln(N)$, where N is the number of processors and K is a synchronization constant \citep{gove_multicore_2010}. This means that at a particular process count, the performance gain over serial code will start decreasing. Minimization of Equation (\ref{eq:runtime}) gives:

			\begin{equation}\label{eq:ncpus}
				N = \frac{T_P}{K}
			\end{equation}

This means that the value of N for which the parallel code scales is directly proportional to the fraction of code that is parallel and inversely proportional to synchronization. In other words, by keeping N constant, one can achieve better performance by either increasing the fraction of parallel code or decreasing the synchronization time, or both.

We have used multiprocessing instead of multithreading to achieve parallelism. There is a very basic difference between threads and processes. \textit{Threads} are code segments that can be scheduled by the operating system. On single processor machines, the operating system gives the illusion of running multiple threads in parallel but in actuality it switches between the threads quickly (time division multiplexing). But in the case of multicore machines, threads run simultaneously on separate cores. Multiple processes are different from multiple threads in the sense that they have separate memory and state from the master process that invokes them (multiple threads use the same state and memory).

The most popular languages for parallel computing are C, C++ and FORTRAN. MPI as well as OpenMP protocols have been  developed for these three languages. But wrappers or software implementations do exist to support interpreted languages like Python, Perl, Java etc. As mentioned in the introduction, the target audience of this paper is astronomers using Python as their language of choice, and/or IRAF. Although much more optimized and faster-executing parallel code can be written in FORTRAN, C or any other compiled language, the main objective here is to optimize astronomer's time. The speed gain with compiled code comes at the cost of longer development time. The secondary objective is \textit{code reuse} i.e. using the existing Python code and/or IRAF tasks to parallelize the problem (wherever it is feasible).

\section{Python Multiprocessing}
\label{python_multiprocessing}
Python supports both multi-threading and multi-processing programming. The threads in Python are managed by the host operating system i.e. scheduling and switching of the threads is done by the operating system and not by the Python interpreter. Python has a mechanism called the Global Interpreter Lock (GIL) that generally prevents more than one thread running simultaneously, even if multiple cores or processors are available \citep{python_software_foundation_python/c_2012}. This results in only one thread having exclusive access to the interpreter resources, or in other words resources are ``locked'' by the executing thread. The running thread releases the GIL for either I/O operations or during interpreter periodic checks (by default after every 100 interpreter ticks or bytecode instructions) \citep{beazley_python_2006}. The waiting threads can run briefly during this period. This unfortunately affects the performance of multi-threaded applications and for CPU bound tasks, the execution time may be actually higher than serial execution. The performance deteriorates further on the multicore machines as the Python interpreter wants to run a single thread at a time whereas the operating system will schedule the threads simultaneously on all the available processor cores.

This is only true for the CPython implementation; PyPy\footnote{Just-in-time (JIT) compilation implementation}, Jython\footnote{Python implementation for Java virtual machine}, and IronPython\footnote{Python implementation for .NET framework} do not prevent running multiple threads simultaneous on multiple processor cores. Jython and IronPython use an  underlying threading model implemented in their virtual machines. But the default Python implementation on most operating systems is CPython and therefore the rest of this paper assumes the CPython implementation. In addition, many of the scientific tools are only available for CPython. Which is another reason why not many in scientific programming use these alternative python distributions.

A better option to get parallel concurrency is to use Python's native \textit{multiprocessing} module. Other standalone modules for parallelizing on shared memory machines include Parallel Python, pyLinda, and pyPastSet. Python multiprocessing and Parallel Python can also be used on a cluster of machines. 

Another parallelizing option for distributed memory machines is message passing. A Python MPI implementation or wrappers (eg. PyCSP \citep{Vinter09pycsprevisited}, mpi4py \citep{dalcin_2008}, pupyMPI  \citep{bromer_2011}, and Pypar \citep{nielsen_2003}) can be used for this purpose. Parallelization can also be achieved by vectorizing the tasks using NumPy. Vectorization is an efficient and optimized way of replacing explicit iterative loops from the Python code. However, not all the operations can be parallelized in Numpy/SciPy \footnote{Refer to http://www.scipy.org/ParallelProgramming for more details}.

We have used the native \textit{multiprocessing} module and the standalone \textit{Parallel Python} module to achieve parallelization. For comparison purposes, we have implemented parallel code for four astronomy routines. 

The \textit{multiprocessing} module is part of Python 2.6 and onwards, and backports exist for versions 2.4 and 2.5\footnote{More details on http://pypi.python.org/pypi/multiprocessing}. \textit{Multiprocessing} can also be used with a cluster of computers (using the multiprocessing \textit{Manager} object), although the implementation is not trivial. This paper deals only with shared memory or symmetric multiprocessing (SMP) machines, although detailed information about computer clusters for astronomical data processing can be found elsewhere\footnote{One such document can be found on our website - http://astro.nuigalway.ie/staff/navtejs}.

A few approaches exist in the Python \textit{multiprocessing} library to distribute the workload in parallel. In this paper we will be considering two main approaches - a pool of processes created by the \textit{Pool} class, and individual processes spawned by the \textit{Process} class. The \textit{Parallel Python} module uses inter-process communication (IPC) and dynamic load balancing to execute processes in parallel. Code implementation using these three approaches is discussed in detail in the following sub-sections. 

\subsection{The Pool/Map Approach} 
Out of the two \textit{multiprocessing} approaches, this is the simplest to implement. The \textit{Pool/Map} approach spawns a pool of worker processes and returns a list of results. In Python functional programming, a function can be applied to every item iterable using the built-in \textit{map} function. For example, instead of running an iterative loop, the \textit{map} function can be used:

\begin{center}
	\begin{lstlisting}[label=lst:iter,caption=Example Python iterative function]
		# Iterative function
		def worker( indata ):
			...
			return result
		
		# Input dataset divided into chunks
		lst = [ in1, in2, ... ]
		
		# Loop over lst items and append results
		results = []
		for item in lst:
			results.append( worker( item ) )
	\end{lstlisting}
\end{center}

\begin{center}
	\begin{lstlisting}[label=lst:map,caption=Python map function]
		# Iterative function
		def worker( indata ):
			...
			return result
		
		# Input dataset divided into chunks
		lst = [ in1, in2, ... ]	
		
		# Iteratively run func in serial mode
		results = map( worker, lst )
	\end{lstlisting}
\end{center}

The \textit{map} function is extended to the \textit{multiprocessing} module and can be used with the \textit{Pool} class to execute worker processes in parallel, as depicted in Listing \ref{lst:pool_map}.

\begin{center}
	\begin{lstlisting}[label=lst:pool_map,caption=Pool/Map multiprocessing implementation]
		# Import multiprocessing module
		import multiprocessing as mp
		
		# Get number of processors on the machine
		# or manually enter required number of processes
		ncpus = mp.count_cpus()
		
		# Define pool of ncpus worker processes
		pool = mp.Pool( ncpus )
		
		# Start ncpus pool of worker processes in parallel
		# output is appended to results python list
		results = pool.map( worker, lst )
	\end{lstlisting}
\end{center}

The \textit{import} command includes the \textit{multiprocessing} module in the routine, the \textit{count\_cpus} method gets the number of processors or cores on the machine, the \textit{Pool} class creates a pool of \textit{ncpus} processes, and \textit{Pool's map} method iterates over the input element list in parallel, and maps each element to the \textit{worker} function. The number of worker processes spawned can be more than the number of cores on the machine but as we will see in Section \ref{examples}, best performance is achieved when the number of processes is equal to the number of physical processor cores or the total number of concurrent threads.
	
\subsection{The Process/Queue Approach}
The Pool/Map approach allows only one argument as an input parameter to the calling function. There are two ways to send multiple arguments: pack arguments in a python list or a tuple, or use the \textit{process} class in conjunction with a \textit{queue} or \textit{pipe}. Although a process can be used without queues and pipes, it is good programming practice to use them\footnote{Python documentation suggests to avoid synchronization locks and instead use the queue or the pipe (http://docs.python.org/library/multiprocessing.html\#programming-guidelines/)}. Two FIFO (First In, First Out) queues are created - one for sending input data elements and another for receiving output data. Parallel worker processes are started using the \textit{Process} class and smaller chunks of input data are put on the send queue for processing. Each worker process picks the next data chunk in the queue after processing the previous chunk. The output result is put on the receive queue, and then read at the end for post-processing.

An example Python code listing for the Process/Queue approach is shown below:

	\begin{lstlisting}[label=lst:process,caption=Process/Queue multiprocessing implementation]
		# Import multiprocessing module
		import multiprocessing as mp
		
		# Worker function
		# iter is standard python built-in function
		def worker( s_q, r_q, arg1, arg2, ... ):
			for value in iter( s_q.get, `STOP` ):
				...
				r_q.put( result )
		
		# Get number of cores on the machine
		ncpus = mp.count_cpus()
		
		# Create send and receive queues
		send_q = mp.Queue()
		recv_q = mp.Queue()
		
		# Start ncpus number of processes
		for i in range( ncpus ):
			mp.Process( target = worker, args = ( send_q, recv_q, arg1, arg2, ... ) )
			
		# Put input data chunks on send queue
		# indata is python list of input data set,
		# which is already divided into smaller chunks
		for chunk in indata:
			send_q.put( chunk )
			
		# Get output from receive queue
		results = []
		for i in range( len( indata ) ):
			results.append( recv_q.get() )
			
		# Stop all the running processes
		for i in  range( ncpus ):
			send_q.put( `STOP` )
	\end{lstlisting}

The code is self-explanatory but the main point to notice is the code related to stopping the running processes. \textit{STOP} or any other value can be put in the send queue to stop the processes, as the worker function reads from the queue until it encounters \textit{STOP}. We have found that in certain conditions, the Process/Queue approach performs better than the Pool/Map approach, as shown in Section \ref{examples}.

\subsection{Parallel Python Approach}
Parallel Python is an open source cross-platform module for parallelizing python code. It provides dynamic computation resource allocation as well as dynamic load balancing at runtime. In addition to executing programs in parallel on Symmetric multiprocessing (SMP) machines, it can also be used on clusters of heterogeneous multi-platform machines \citep{vanovschi_parallel_2013}.

Processes in Parallel Python run under a job server. The job server is started locally (or remotely if running on a cluster of machines) with the desired number of processes (ideally equal to the number of processor cores). A very basic code template is shown below:

	\begin{lstlisting}[label=lst:pp,caption=Parallel Python implementation]
		# Import Parallel Python module
		import pp
		
		# Parallel Python worker function
		def worker( indata ):
			...		
			return result

		# Create an empty tuple
		ppservers = ()
		
		# Either manually set the number of processes or
		# default to the number of cores on the machine
		if ncpus:
			job_server = pp.Server( int(ncpus), ppservers = ppservers )
		else:
			job_server = pp.Server( ppservers = ppservers )
		
		# Divide data into smaller chunks for better 
		# performance (based on the scheduler type)
		chunks = getchunks( infile, job_server.get_ncpus(), scheduler )
		
		# Start the worker processes in parallel
		jobs = []
		for value in chunks:
			indata = ( arg1, arg2, ...  )
			jobs.append( job_server.submit( worker, (indata,), (func1,func2,...), (mod1,mod2,...) ) )
			
		# Append the results   
		results = []
		for job in jobs:
			results.append( job())
	\end{lstlisting}

The number of processes (\textit{ncpus}) is passed to the program as user input, or defaulted to the number of cores on the machine. The input data is divided into smaller chunks for better load balancing. The \textit{Parallel Python} job server starts the worker processes for parallel execution. At the end of the execution, results are retrieved from the processes. The parameters to the job server's submit method are the worker function, its arguments, and any other function or module used by the worker function.

\section {Benchmarking}
\label{examples}

To benchmark the parallel processing approaches described in the previous section, three different kinds of astronomical problems, of varied complexity, were parallelized. Three machines of different configuration were used to benchmark the code. The hardware and software configuration of the test machines is listed in Table \ref{tab:machines}.

\begin{table}[h]
\centering
	\def\arraystretch{1.5}
	{\footnotesize
	\begin{tabular}{ | p{1.0cm} | p{2.1cm} | p{1.0cm} | p{2.0cm} | }
		\hline
		\multicolumn{1}{| c}{\textbf{Machine}} & \multicolumn{1}{| c |}{\textbf{Processor}} & \multicolumn{1}{| c |}{\textbf{Memory}}& \multicolumn{1}{| c |}{\textbf{Operating System}} \\
		\hline
		\hline
		Homebuilt &  AMD Phenom II X4 B60 quad core processor @3.4GHz & 4 GB @1066 MHz & Ubuntu 11.10 32-bit. Linux Kernel 3.0.0-16-generic-pae \\
		\hline
		Dell Studio XPS & Intel Core i7 920 quad core processor (8 threads) @2.67GHz & 6 GB @1066 MHz &  Ubuntu 12.10 32-bit. Linux Kernel 3.5.0-22-generic \\
		\hline
		iMac 21.5 inch & Intel Core i5 I5-2400S quad core processor @2.5GHz & 4 GB @1333 MHz & Mac OS X 10.8.1 64-bit. Darwin 12.1.0 \\
		\hline
	\end{tabular}}
	\caption{Hardware and software configuration of the SMP machines used for benchmarking the parallel code.}
	\label{tab:machines}
\end{table}

The homebuilt machine was over-clocked to 3.51 GHz and was running the Ubuntu 11.10 operating system (OS). The Dell Studio XPS was running the same OS but as a guest OS in a VirtualBox\footnote{VirtualBox is open source virtualization software. More details on https://www.virtualbox.org/} virtual machine on a Windows 7 host. Out of the 6 GB of RAM on the Dell machine, 2 GB was allocated to the guest OS. The iMac was running Mac OS X Mountain Lion. For astronomical data and image processing, existing routines from the ESO Scisoft 7.6\footnote{ESO Scisoft software package is a collection of astronomical utilities distributed by European Southern Observatory} software package were used. This version of Scisoft is using Python 2.5 and therefore the python \textit{multiprocessing} backport was separately installed. The latest version of Scisoft (version 7.7 released in March 2012) has been upgraded to Python 2.7 and therefore does not require separate installation of the \textit{multiprocessing} module. The iMac machine was using the latest version of Scisoft.

All of our benchmarking examples concern a class of parallel work flow known as \textit{Embarrassingly Parallel} problems. In simple terms this means that the problem can be easily broken down into components to be run in parallel. The first astronomy routine is a parallel implementation of coordinate transformation of the charge-coupled device (CCD) pixel to sky coordinates (RA and DEC), and vice-versa, for Hubble Space Telescope (HST) images, for a large number of input values. The second routine parallelized a Monte Carlo completeness test. The third routine is parallel implementation of sub-sampled deconvolution of HST images with a spatially varying point spread function (PSF). These routines are described in the following sub-sections. They are freely available and can be downloaded from our website\footnote{http://astro.nuigalway.ie/staff/navtejs}. 

\subsection{PIX2SKY and SKY2PIX}
\label{pix2sky}
The STSDAS\footnote{Space Telescope Science Data Analysis System} package for IRAF includes the routines \textit{xy2rd} and \textit{rd2xy} to transform HST CCD image pixel coordinates to sky coordinates, and vice-versa. These tasks can only process one coordinate transformation at a time. Running them serially, or even in parallel, to process hundreds to thousands of transformations (e.g. star coordinate transformations in massive star clusters) is not efficient, as it will be performing the same expensive FITS header keyword reads on each call. Non-IRAF routines to transform lists of coordinates do exist, but they default to serial processing. Two such routines (\textit{xy2sky} and \textit{sky2xy}, implemented in C) are part of the WCSTools\footnote{More details on http://tdc-www.harvard.edu/wcstools/} software. A pure Python implementation exists in the \textit{pywcs} module, which is part of the \textit{astropy} package\footnote{Refer to http://www.astropy.org/ for more details}.

As these IRAF routines are not lengthy, they can easily be re-written for multicore machines to process a large input dataset in parallel. We have implemented two tasks - PIX2SKY (based on \textit{xy2rd}) and SKY2PIX (based on \textit{rd2xy}) -  in Python, using both the \textit{multiprocessing} and \textit{Parallel Python} modules. These routines read input files with either [X,Y] pixel coordinates (PIX2SKY routine) or [RA, DEC] values (SKY2PIX routine), and output transformed coordinates. We have used the PyFITS\footnote{STScI's python module for working with FITS files} module from the Space Telescope Science Institute (STScI) for handling FITS images.
 
\begin{figure*}[!ht]
	\centering
		\includegraphics[width=0.76\linewidth]{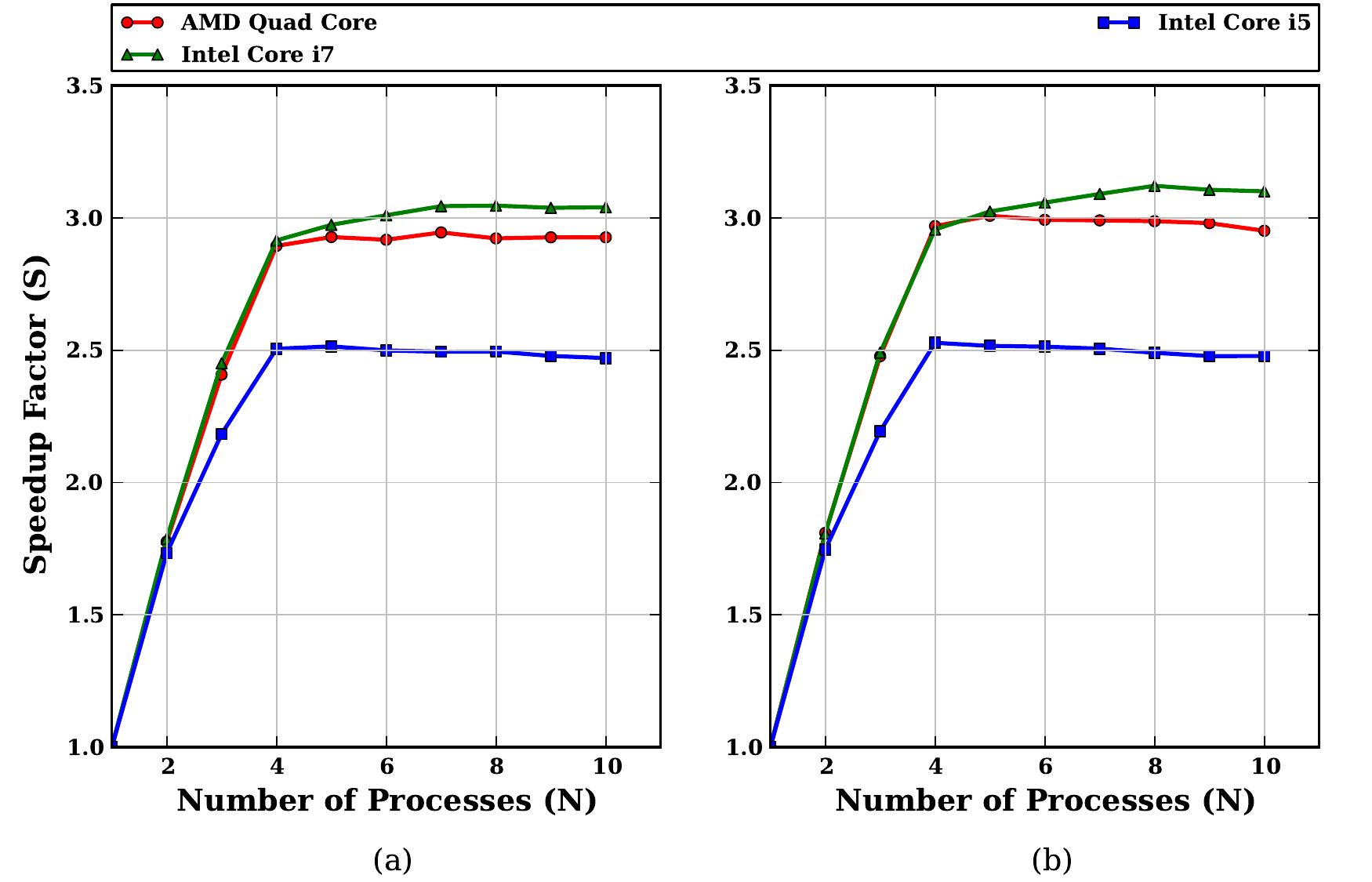}
		\caption{Coordinate transformation benchmark, using the Parallel Python approach with a guided scheduler. (a) PIX2SKY  and (b) SKY2PIX speedup on the three test machines. Data points are the average of 50 runs. As expected, maximum speedup occurs when the number of processes is equal to the number of physical cores on the machine. Of course, load balancing and the serial to parallel code fraction also influence the speedup.}
		\label{fig:fig1}
\end{figure*}

The speedup factor (as defined in Section \ref{pdp}) for PIX2SKY and SKY2PIX is plotted against the number of processes in Figure \ref{fig:fig1}. For this benchmarking, one million input coordinates were fed into the \textit{Parallel Python} based transformation modules with guided scheduling.

Best performance was achieved when the number of processes was equal to the number of physical cores on the machine. The Intel Core i7 based machine showed better speedup than the AMD Phenom based machine. This is because Intel Core i7 processors use hyper-threading technology\footnote{Proprietary technology of Intel Corporation to allow multiple threads to run on each core} and have 2 threads per core. Theoretically, we would have expected close to eight-fold speedup on the Intel Core i7 quad-core processor (with hyper-threading), but bottlenecks like I/O operations restrict higher performance gains (keeping the fraction of code running in parallel constant). As the number of processes increases beyond the number of cores, the speedup factor is almost flat. This is a result of dynamic load balancing in Parallel Python, which is not the case for the Process/Queue approach (as we will see in Section \ref{mmp}).  

Another interesting point regarding Figure \ref{fig:fig1} is the somewhat inferior performance of Core i5 processor compared to the other two quad core processors. We could not do a direct comparison between the Core i5 processor and other two processors as by default \textit{Turbo Boost}\footnote{Proprietary technology of Intel Corporation to run processors above their base frequency} and \textit{Speed Step}\footnote{Refer to http://en.wikipedia.org/wiki/SpeedStep for more details} are turned on in the iMac machine (and cannot be switched off). For the other two processors - \textit{Speed Step technology} was turned off on the Core i7 processor and \textit{Cool'n'Quiet}\footnote{Refer to http://www.amd.com/us/products/technologies/cool-n-quiet/Pages/cool-n-quiet.aspx for more details.} was switched off on the AMD machine to prevent CPU throttling.

\subsection{Completeness Test}
In more complex or lengthy problems, it makes much more sense to reuse the existing tasks or optimized serial codes, rather than re-writing code as we did in Section \ref{pix2sky}. A good example is a completeness test - the method to determine the detection efficiency of different magnitude stars in images. It is a Monte Carlo class of algorithm, i.e. it involves repeated random execution. The basic idea behind the method is to add artificial stars of varied magnitudes to random positions within the image, and then determine their recovery rate as a function of magnitude. For the test to be statistically relevant, an average of a few hundred random iterations of modified images is taken.

To reuse the existing code, the \textit{addstar} task in the DAOPHOT package of IRAF was used to add artificial stars to the image under question and the \textit{daofind} and \textit{phot} tasks in DAOPHOT were used to detect and match the added stars. One way to achieve parallelism in this case is to run multiple iterations in parallel on the available computing nodes, where each node gets a complete image.

\begin{figure*}[!ht]
	\centering
		\includegraphics[width=0.76\textwidth]{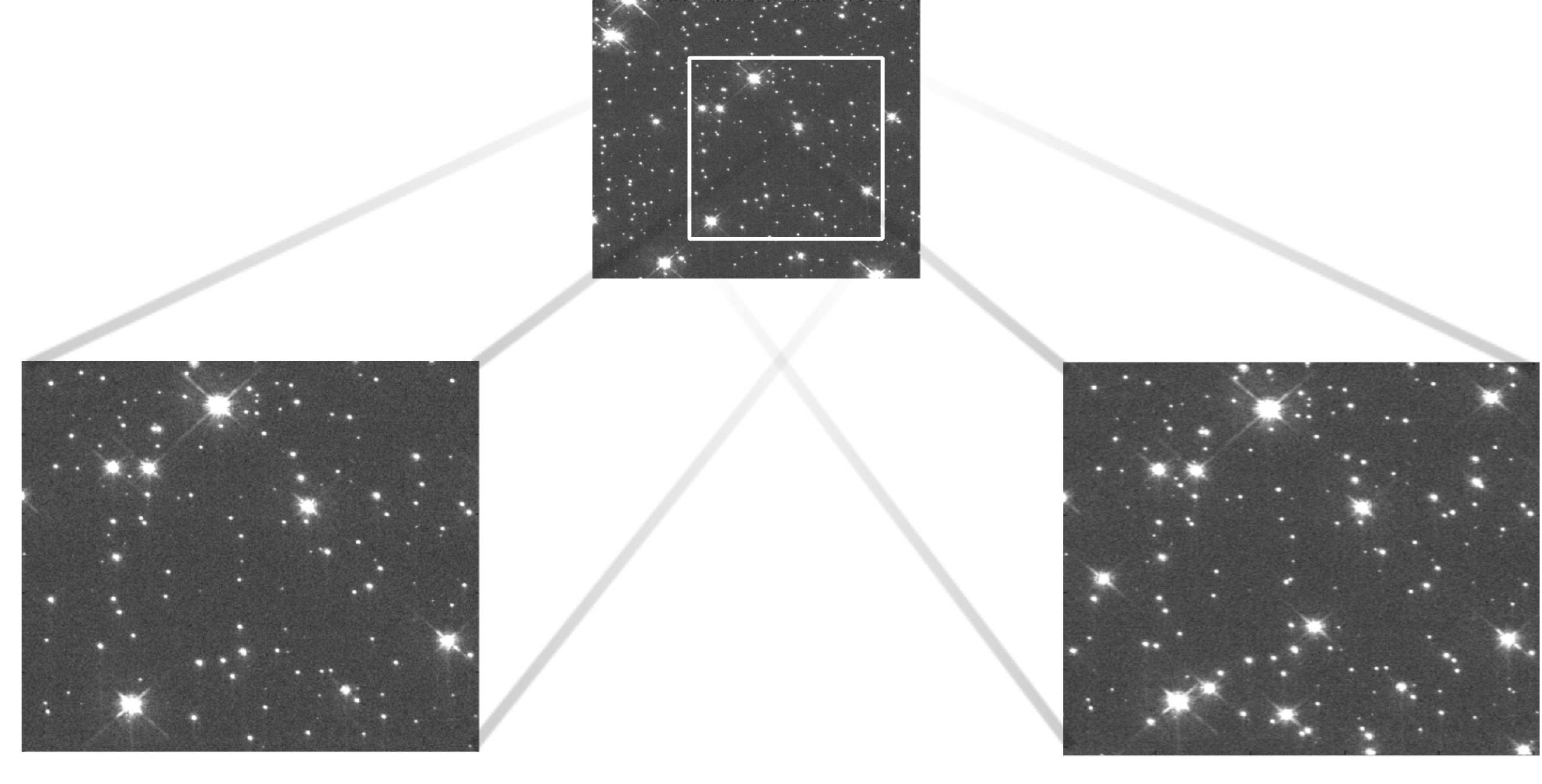}
		\caption{HST WFPC2 co-added image of the globular cluster M71 (NGC 6838) in the F555W filter, used for benchmarking the completeness routine. The test was performed on a 512$\times$512 image section (lower left). This image plus 50 artificial stars between $14^{th}$ and $28^{th}$ magnitude, generated in one of the Monte Carlo iterations, is shown in the lower right.}
		\label{fig:fig2}
\end{figure*}

For the benchmark, we used the reduced and co-added HST WFPC2\footnote{Wide Field Planetary Camera 2} (chip PC1) image of galactic globular star cluster M71 (NGC 6838) in the F555W filter (shown in Figure \ref{fig:fig2}). The completeness test was performed on a 512$\times$512 pixel uncrowded section of the image with $14^{th} - 28^{th}$ magnitude stars in each iteration. Fifty stars (a sufficiently small number to avoid changing the star density appreciably) were added randomly to the image section (shown in Figure \ref{fig:fig2}), and 100 iterations per magnitude interval were run. As in the previous benchmark, we used the Parallel Python approach with guided scheduler.

\begin{figure*}[!ht]
	\begin{minipage}[c]{0.5\linewidth}	
		\subfigure[]{\label{fig:fig3a}\includegraphics[width=1\linewidth]{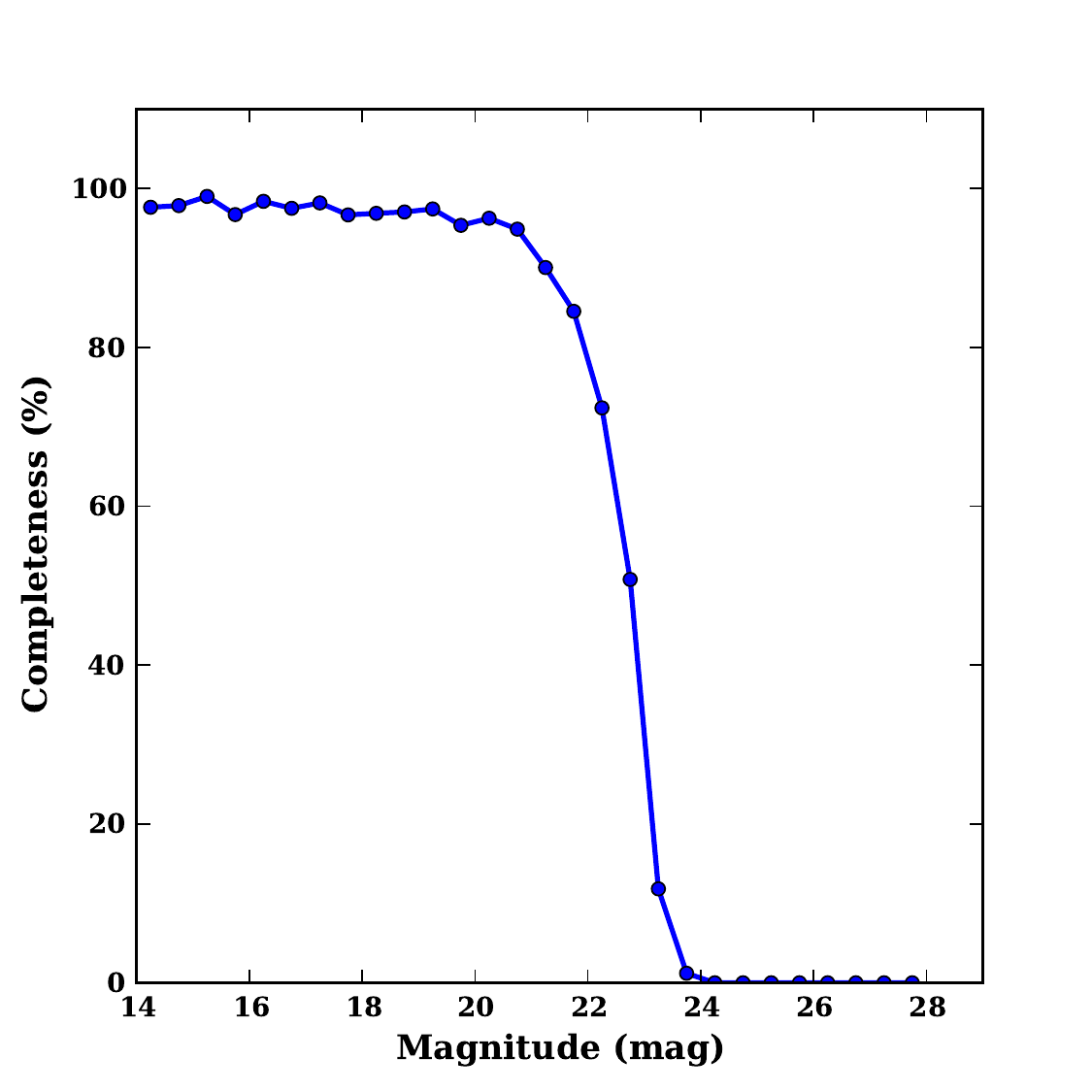}}
	\end{minipage}		
	\begin{minipage}[c]{0.5\linewidth}
		\subfigure[]{\label{fig:fig3b}\includegraphics[width=1\linewidth]{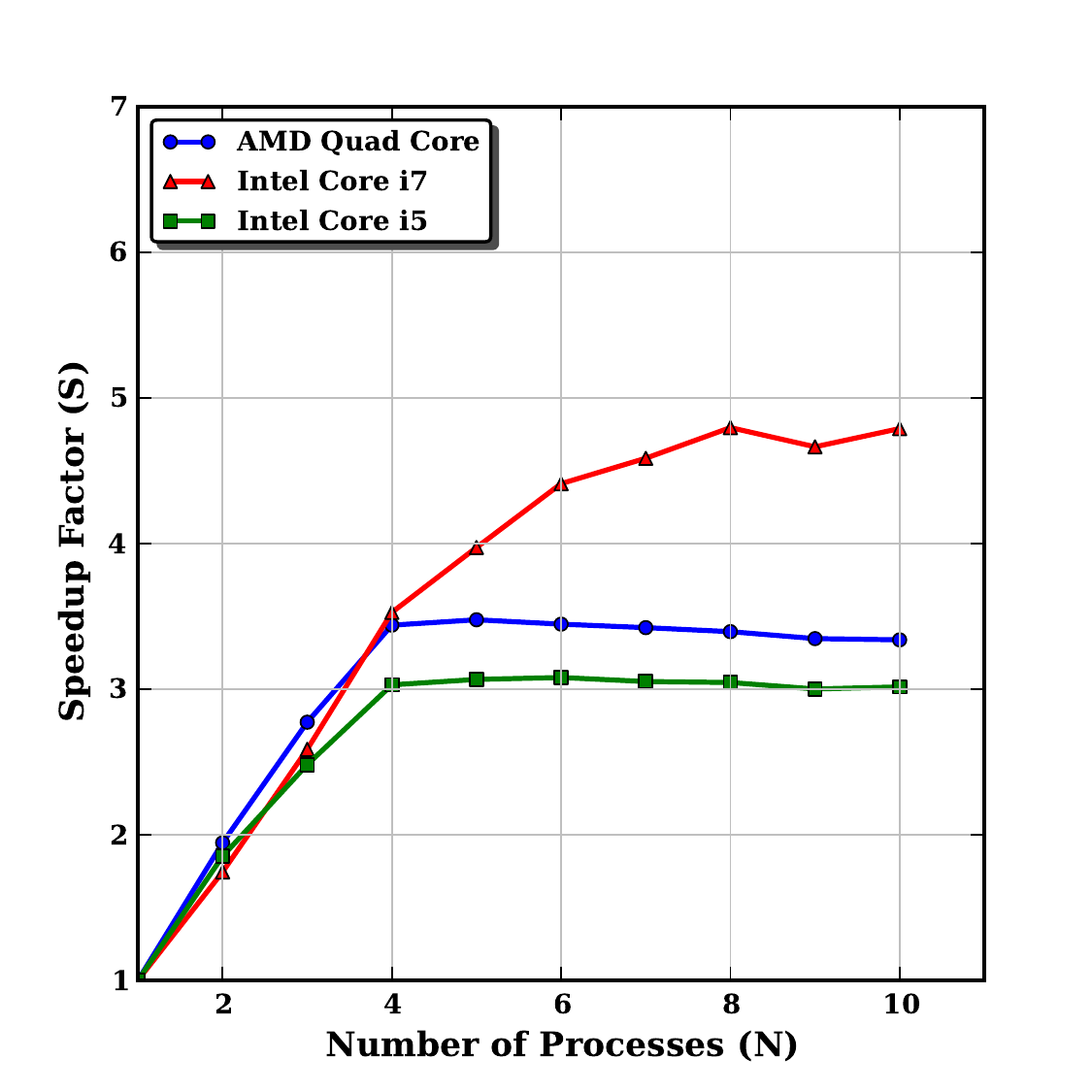}}
	\end{minipage}		
		\caption{(a) Detection completeness plot for the images in Figure \ref{fig:fig2} with artificial star magnitudes varying from 14.0 to 28.0 mag within 0.5 mag intervals; (b) Speedup achieved on the test machines as a function of the number of parallel processes. The Parallel Python approach with guided scheduler was used for benchmarking.}
\end{figure*}

In Figure \ref{fig:fig3a}, we plot the completeness test results, computed by averaging results from all the iterations, with a 0.5 magnitude bin interval. Figure \ref{fig:fig3b} shows that maximum speedup is achieved when the number of processes is equal to the number of physical cores, or the total number of threads (in the case of the Intel Core i7 machine). All three processors show close to 2$\times$ speedup with 2 spawned processes. After four processes, speedup flattens out for the AMD and Core i5 processors whereas it keeps on increasing for the Core i7 machine (although not as steeply as from 1 to 4 processes). This is explained by Equation \ref{eq:ncpus} - the number of processes is scaled on the parallel code fraction and synchronization time.

\subsection{Parallel Sub-Sampled Deconvolution}
Image deconvolution is an embarrassingly parallel problem, as image sections can be deconvolved in parallel and combined at the end. In addition to being under-sampled, HST detector point spread functions are spatially varying. To recover resolution lost to aberration and poor sampling in HST images, \citet{butler_deconvolution_2000} proposed an innovative sub-sampled deconvolution technique. We have implemented a parallel version of this technique.

\begin{figure*}[!ht]
\begin{center}
	\subfigure[]{\label{fig:fig4a}\includegraphics[width=0.52\linewidth]{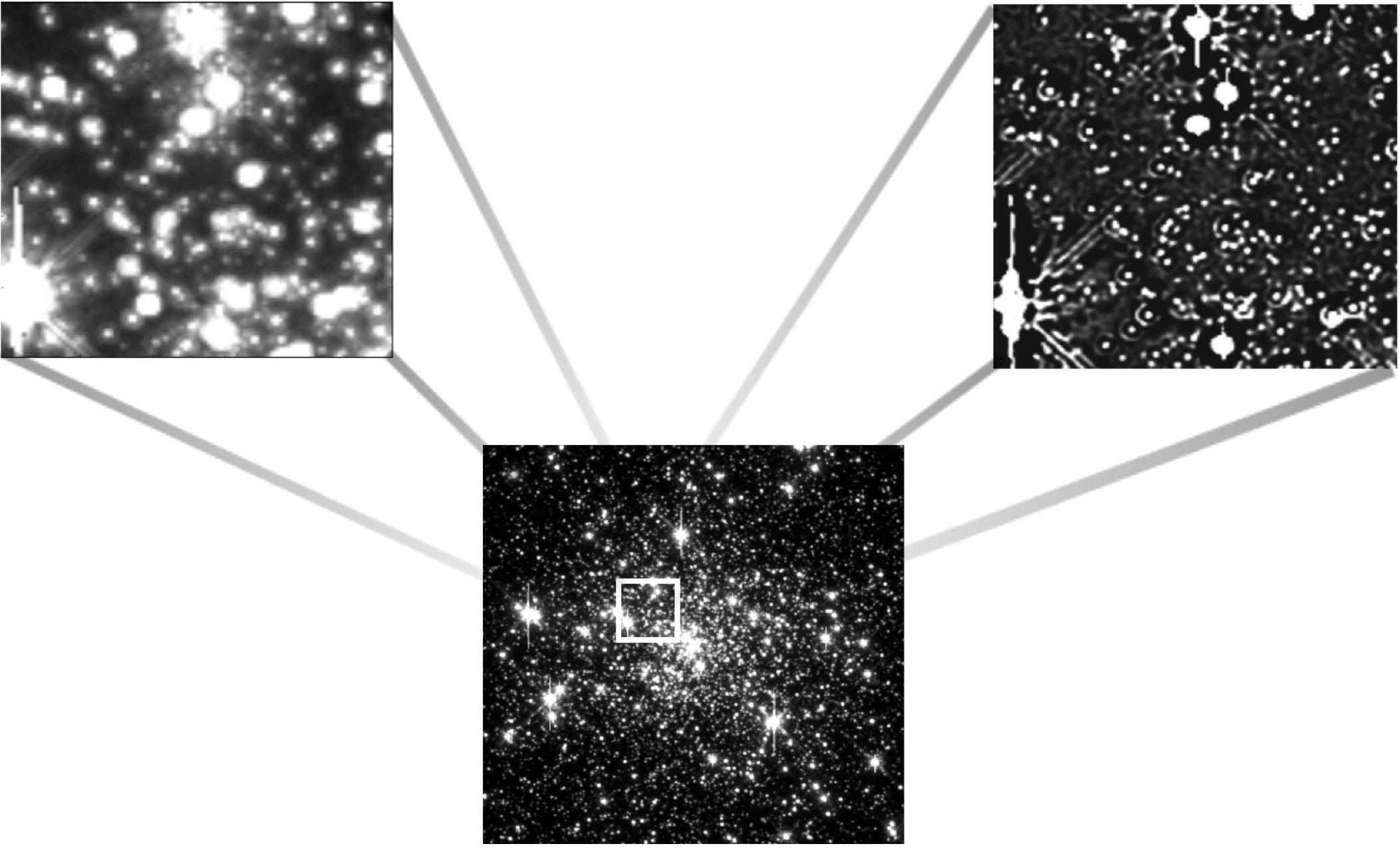}}
	\hspace{20pt}
	\subfigure[]{\label{fig:fig4b}\includegraphics[width=0.43\linewidth]{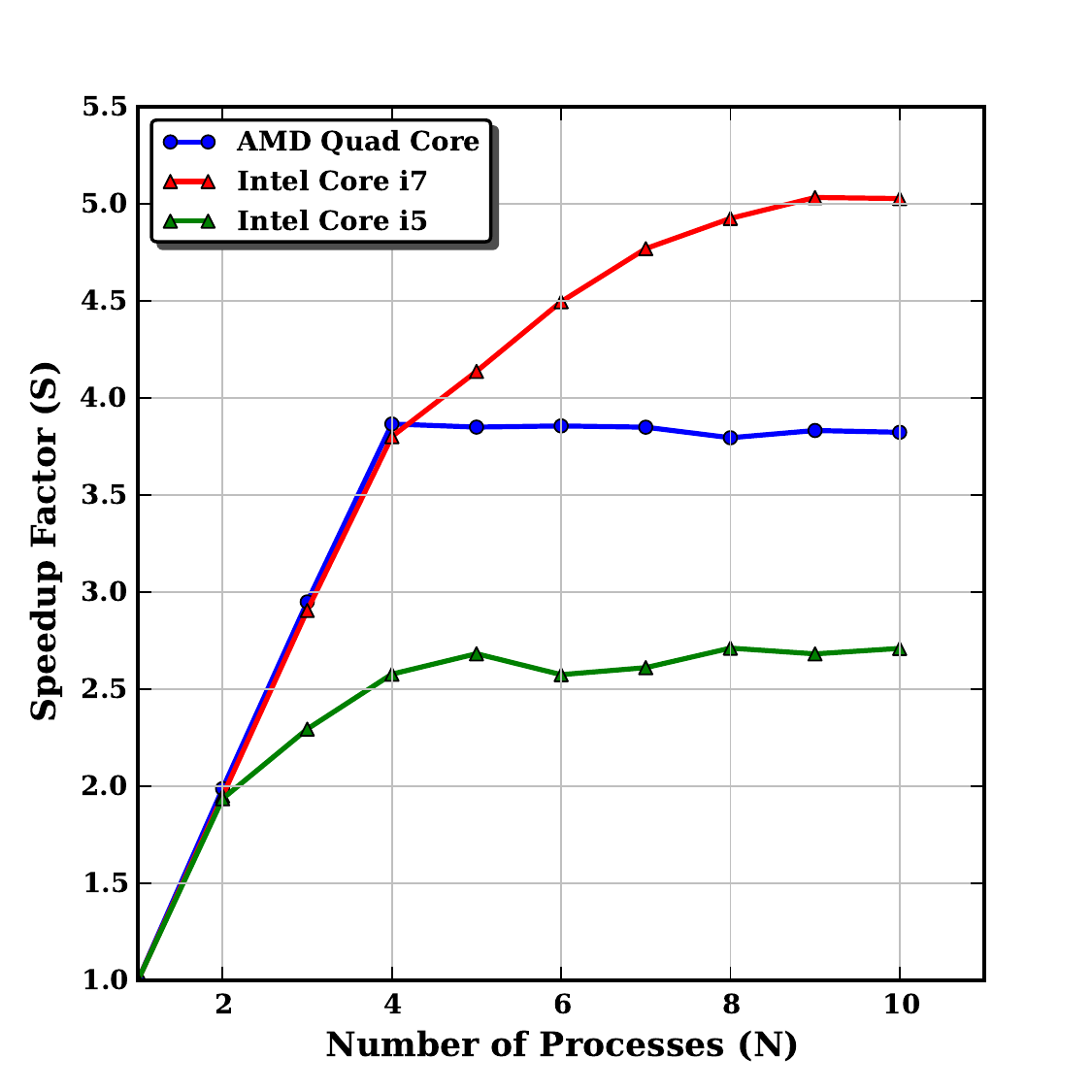}}
\end{center}	
	\caption{(a) HST WFPC2 PC1 coadded image of globular cluster NGC 6293 in the F814W filter. A normal sampled $4.6\arcsec \times 4.6\arcsec$ image section is shown in the upper left hand corner, and the 2$\times$ sub-sampled deconvolved section in the upper right hand corner; (b) Sub-sampled deconvolution benchmark. The AMD quad core performance levels out after four processes, whereas the Intel quad core achieves a speedup factor of almost 5 for eight processes. The Parallel Python approach with guided scheduler was used for benchmarking.}
\end{figure*}

To deconvolve HST images with a spatially varying PSF, we wrote a parallelized version of the sub-sampled image deconvolution algorithm. The Maximum entropy method (MEM) implementation in the STSDAS package was used for deconvolution. As it only uses a spatially invariant PSF, highly overlapping 256$\times$256 sub-images were deconvolved with the appropriate PSF for that position on the CCD. Deconvolved sub-images were reassembled to generate the final sub-sampled deconvolved image.

For benchmarking, we used a coadded HST WFPC2 PC1 chip image of the globular cluster NGC 6293. A spatially varying analytical PSF model was generated from the TinyTim\footnote{TinyTim is a point spread function modeling tool for HST instruments} PSF grid. The Parallel Python module with guided scheduler was used for benchmarking. The central region of NGC 6293, along with a high resolution section of the normal sampled image and sub-sampled deconvolved image, are shown in Figure \ref{fig:fig4a}. The speedup factor levels out close to 4 for the AMD machine, whereas it levels out at 5 for the Intel Core i7 machine (see Figure \ref{fig:fig4b}). In general, embarrassingly parallel problems provide the best speedup as a higher fraction of code is running in parallel.

\section{Discussion}
\label{discuss}

\subsection{Ease of Implementation}
As shown in Section \ref{python_multiprocessing}, parallelizing existing python serial code can be pretty straightforward for problems that can be easily broken down into smaller parts. This is especially true if the \textit{multiprocessing} Pool/Map approach is used in existing functional python code. Python includes excellent code profilers - \textit{cProfile} and \textit{Profile}. These, along with the \textit{pstats} module, can be used to optimize the parallel code as well as the serial code.

Astronomers using IRAF scripts now have another reason to move to PyRAF, and the python scripting language in general. The PyRAF command line interface allows execution of IRAF CL scripts, and also gives full flexibility and the power of the python language. STScI has a very useful PyRAF introductory tutorial\footnote{http://stsdas.stsci.edu/pyraf/doc.old/pyraf\_tutorial/} on their website. Another reason for using python is that it is an open source language, supported by community development, with most of the modules available for free. Python modules NumPy and SciPy comprise an extensive collection of numerical and scientific functions\footnote{Numerical Python (NumPy) and Scientific Python (SciPy) are non-standard Python modules}.

IRAF saves the task parameter values in global parameter files. This could lead to deadlock if same two tasks try to gain write access to the parameter file. This can be avoided in PyRAF by explicitly specifying all the parameters in the task, thus avoiding accessing the global parameter file.

\subsection{Load Balancing}
Load balancing in parallel computing is the distribution of the workload on computing nodes to optimize performance. It is an important issue to deal with while designing a parallel application. On multicore machines, all the cores will not perform identically as other activities and tasks are running concurrently. A parallel program will run as fast as the slowest core. Thus, the performance of a parallel program will be dictated by the slowest processor core. Therefore, efficient load balancing of the workload on the processor cores is very important. On shared memory machines, the OpenMP protocol has implemented four types of schedulers or load balancing routines - \textit{static}, \textit{guided}, \textit{dynamic} and \textit{runtime} \citep{chandra_parallel_2001}. On the same lines, we have implemented \textit{static} and \textit{guided} scheduler routines to slice large input datasets into smaller chunks and distribute them on the computing nodes. In the static scheduler, equal chunks of data are distributed on each computing node. In the guided scheduler, instead of distributing equal chunks in one go, the dataset is divided in much smaller chunks to provide better performance. These two schedulers were used in both the PIX2SKY and SKY2PIX routines. For more efficient load balancing, dynamic or runtime routines can be implemented. 

In normal usage of both PIX2SKY and SKY2PIX, we read hundreds to thousands of input records. Sending large chunks of file data to computing nodes would result in a longer synchronization time. Instead, we send the starting and ending byte values of the file chunks, and actual file reading is done inside the worker process at each node. Example scheduler implementation code is shown below:

	\begin{lstlisting}[label=lst:load,caption=Scheduler implementation code]
		def getchunks( infile, ncpus, scheduler = `guided` ):
			# Based on scheduler, divide the input data element
			if scheduler == `static`:
				size = getsize( infile ) / ncpus
			else:
				size = getsize( infile ) / (ncpus * 20)
			
			# Open the input file. Display error and 
			# exit if open fails	
			try:
				ifile = open( infile )
			except:
				print >> sys.stderr, `Error : Not able to open `, infile, `. Exiting.`
				sys.exit( -1 )
			
			# Get starting and ending byte values for each chunk, 
			# taking care to read the full line	
			while 1:
				start = ifile.tell()
				ifile.seek(size, 1)
				s = ifile.readline()
				yield start, ifile.tell() - start
				if not s:
					break
			
			# Close the input file	
			ifile.close()
    	\end{lstlisting}
 
\begin{figure*}[!ht]
	\centering
	\includegraphics[width=0.8\linewidth]{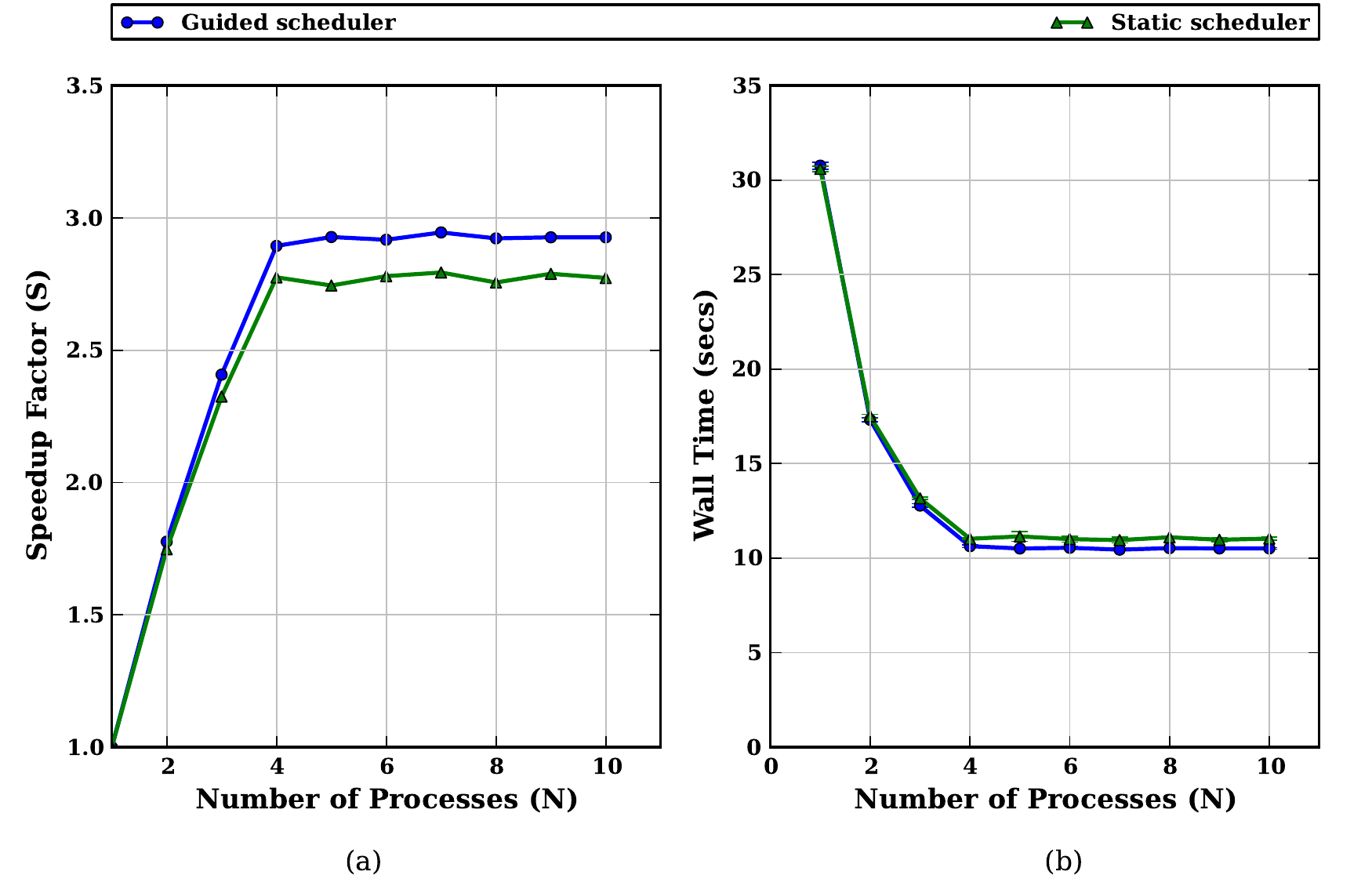}
	\caption{(a) Speedup and (b) Wall time for the PIX2SKY routine with different schedulers. The guided scheduler shows better performance than the static. Wall time is the total execution time for the routine. Data points are averaged over 100 runs. Program wall time on COTS machines depends on the other jobs and programs running. Error bars depict 1 standard deviation. The Parallel Python approach was used.}
	\label{fig:fig5}
\end{figure*}

In the case of the \textit{guided} scheduler, the number of records is divided into a large number (20 times the number of processes) of smaller chunks. The performance of the PIX2SKY routine with one million input records on our AMD quad core machine, with both the \textit{static} and \textit{guided} scheduler, is shown in Figure \ref{fig:fig5}. The Parallel Python approach was used, but we get the same results with the other two approaches also.

The graph clearly shows a better performance for the guided scheduler. Therefore dividing work into smaller fragments is better for optimum performance of parallel programs.

\subsection{Multiprocessing Method Performance}
\label{mmp}
\begin{figure*}[!ht]
	\begin{minipage}[c]{0.5\linewidth}
		\subfigure[PIX2SKY]{\label{fig:fig6a}\includegraphics[width=0.92\linewidth]{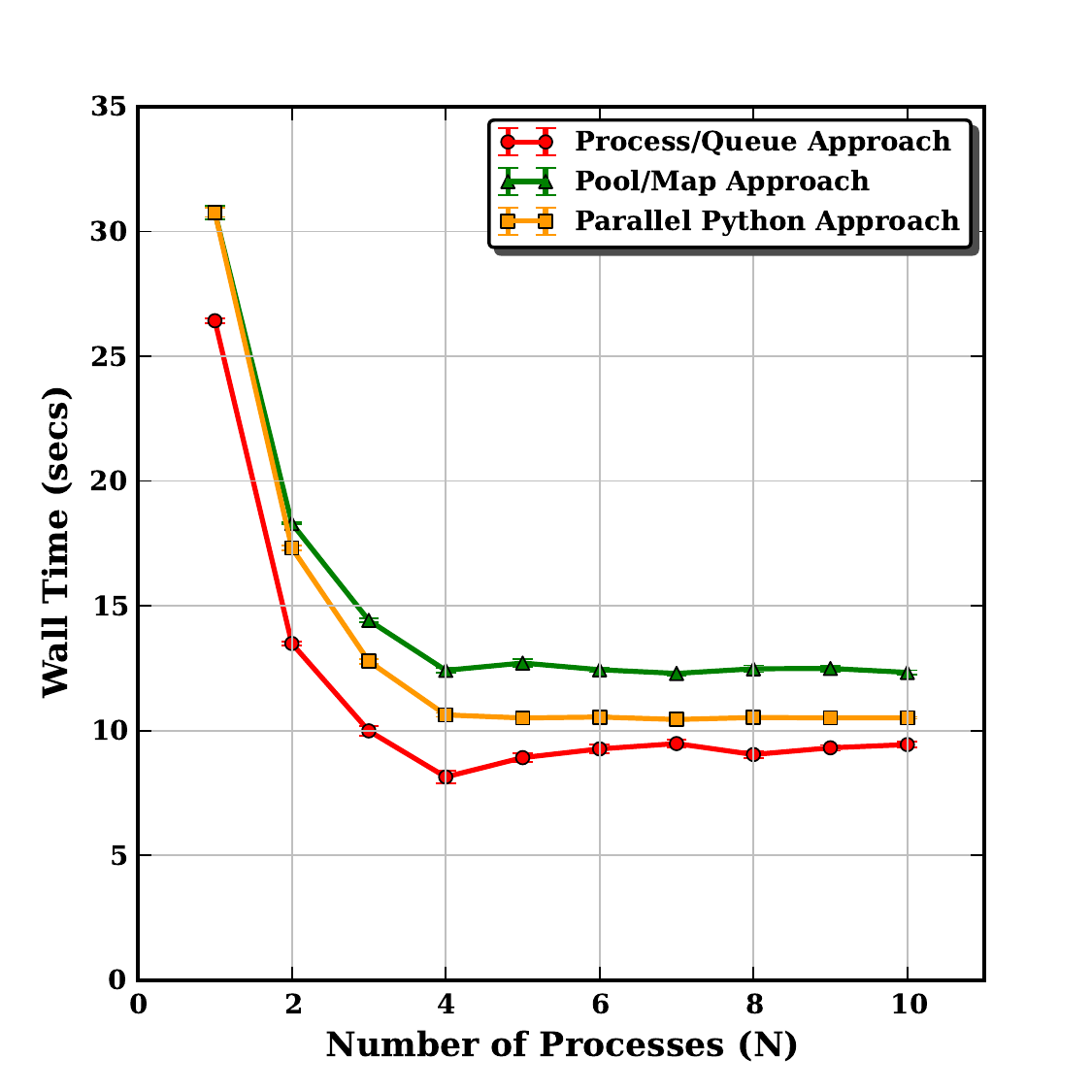}}
	\end{minipage}
	\hfill
	\begin{minipage}[c]{0.5\linewidth}
		\subfigure[SKY2PIX]{\label{fig:fig6b}\includegraphics[width=0.92\linewidth]{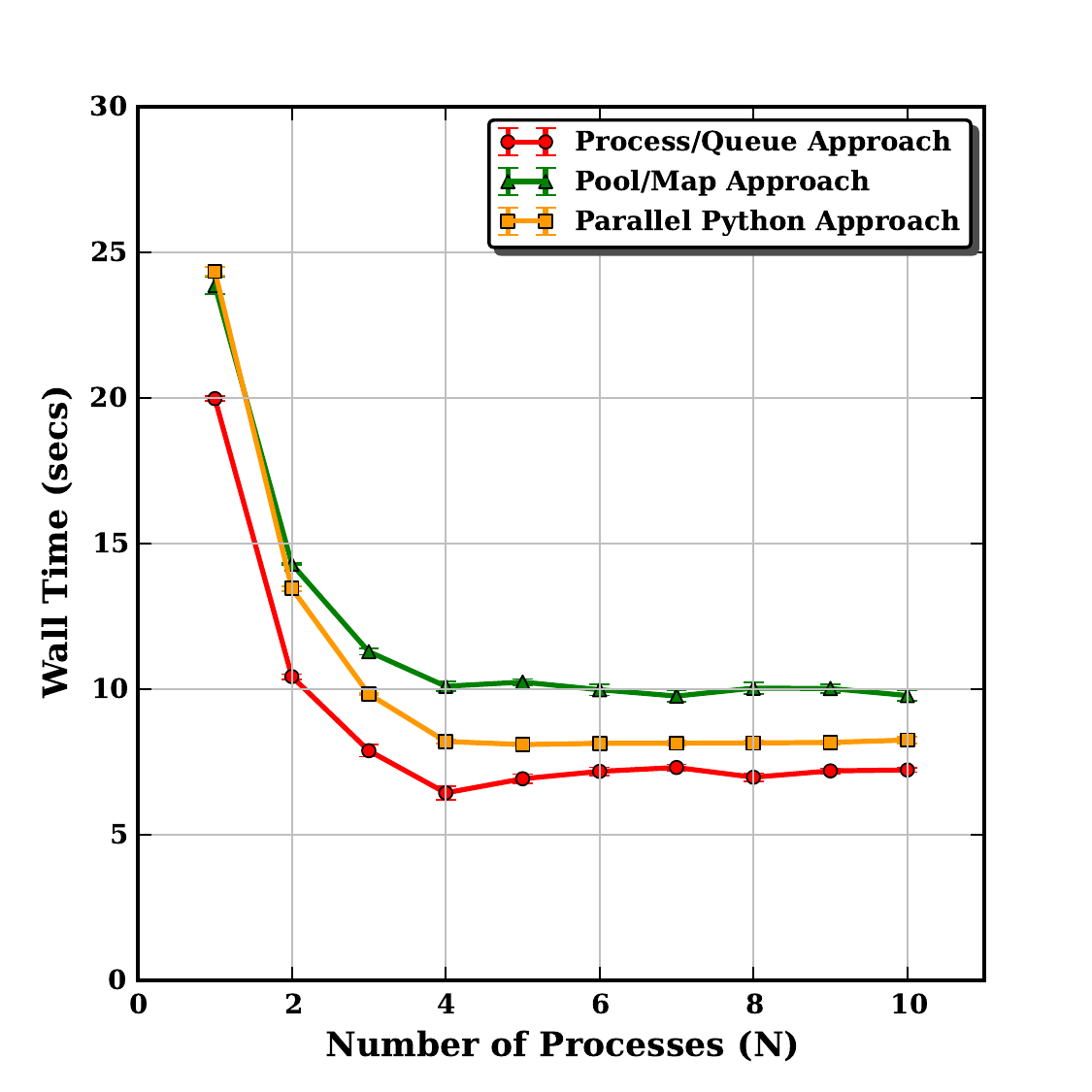}}
	\end{minipage}
	\begin{minipage}[c]{0.5\linewidth}
		\subfigure[Completeness Test]{\label{fig:fig6c}\includegraphics[width=0.92\linewidth]{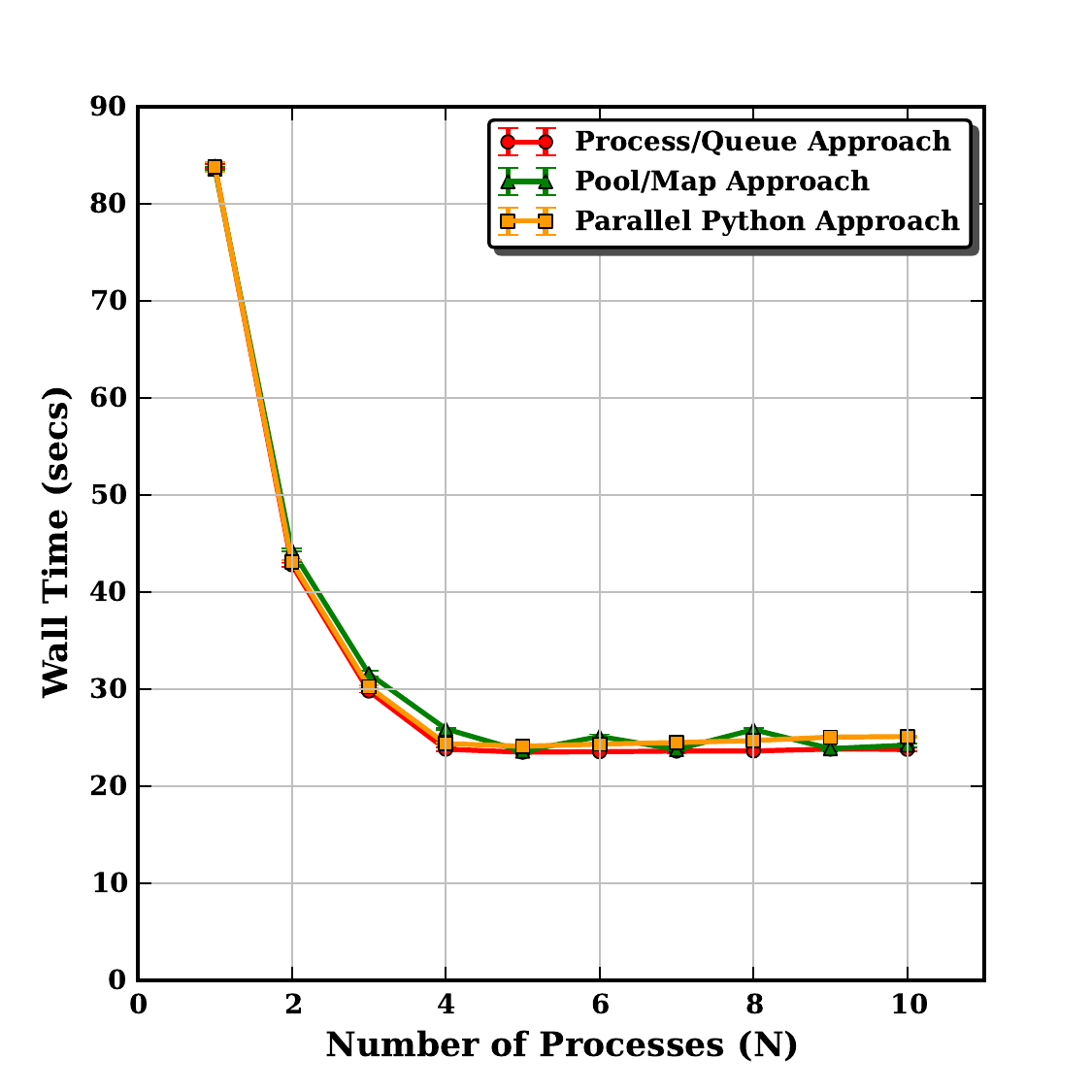}}
	\end{minipage}	
	\begin{minipage}[c]{0.5\linewidth}
		\subfigure[Sub-sampled Deconvolution]{\label{fig:fig6d}\includegraphics[width=0.92\linewidth]{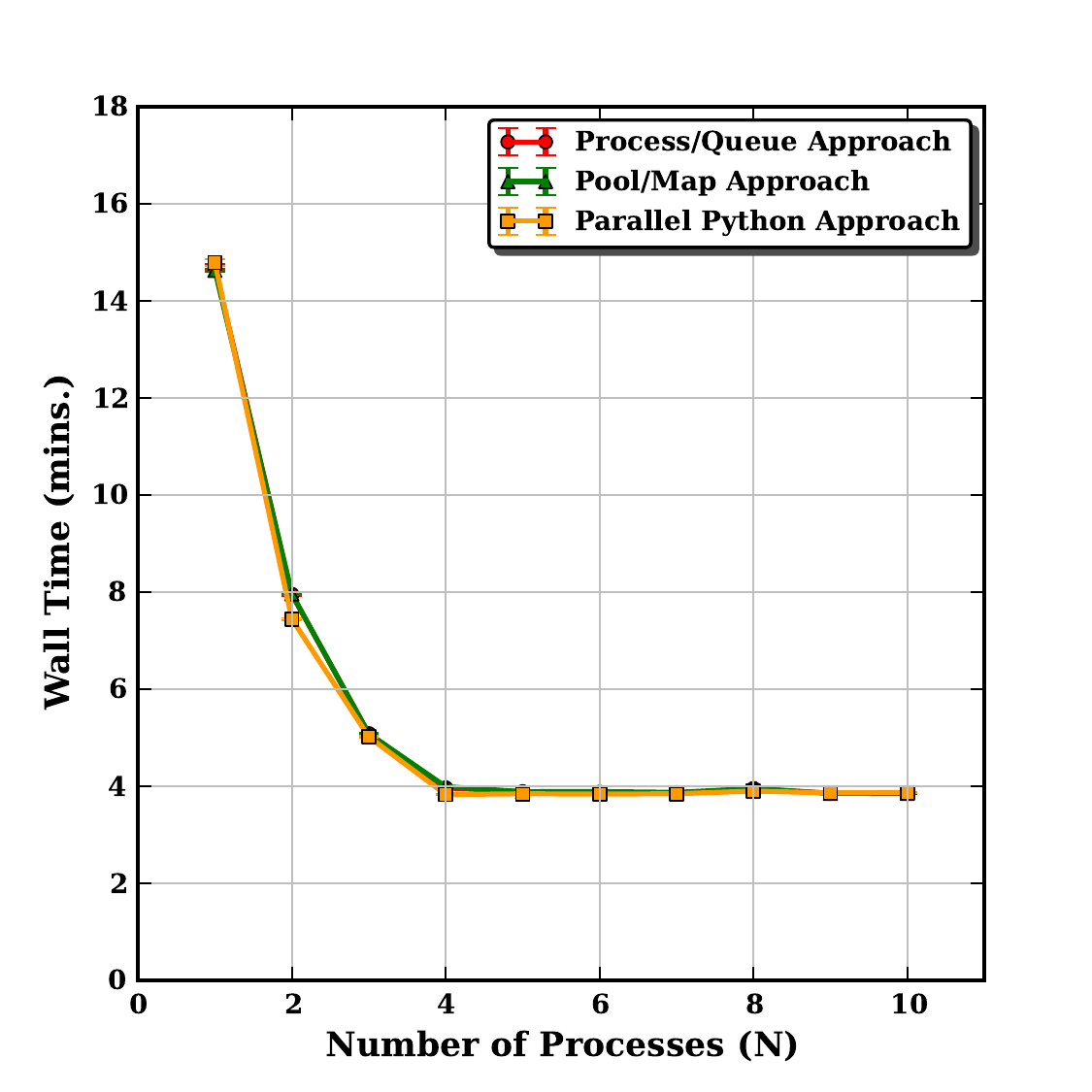}}
	\end{minipage}		
	\caption{Performance of different multiprocessing methods. Panels (a) and (b) depict the performance of our PIX2SKY and SKY2PIX routines on our AMD quad core machine respectively. Panel (c) depicts the performance of our completeness test and panel (d) that of our parallel sub-sampled deconvolution.  Data points are the average of 50 iterations for the PIX2SKY benchmark and 10 iterations for the completeness test and deconvolution. The Parallel Python, Pool and Process approaches show comparable performance for our completeness and sub-sampled deconvolution routines, whereas the Process/Queue approach performs better than both the Parallel Python and Pool/Map approaches for the PIX2SKY and SKY2PIX routines.}
\end{figure*}

To compare the performance of the different multiprocessing methods introduced in Section \ref{python_multiprocessing}, we again took the PIX2SKY and SKY2PIX routines with one million input coordinates to be transformed. Our plots of run time (wall time) versus number of processes for the AMD quad core machine are shown in Figure \ref{fig:fig6a} and Figure \ref{fig:fig6b}.

For both PIX2SKY and SKY2PIX, the Process/Queue approach performs better than both the Pool/Map and Parallel Python approaches. However, Parallel Python scales nicely when the number of processes are more than number of physical cores. Although the Process/Queue performs better, implementing the Pool/Map method is much more straightforward. 

For our routines using existing IRAF tasks - the completeness test and sub-sampled deconvolution - all three approaches have comparable performance. This is because the execution time of these routines is dictated by the performance of the underlying IRAF tasks (see Figure \ref{fig:fig6c} and Figure \ref{fig:fig6d}). 

\subsection{Scalability and Portability}
\begin{figure*}[!ht]
	\centering
		\includegraphics[width=0.6\linewidth]{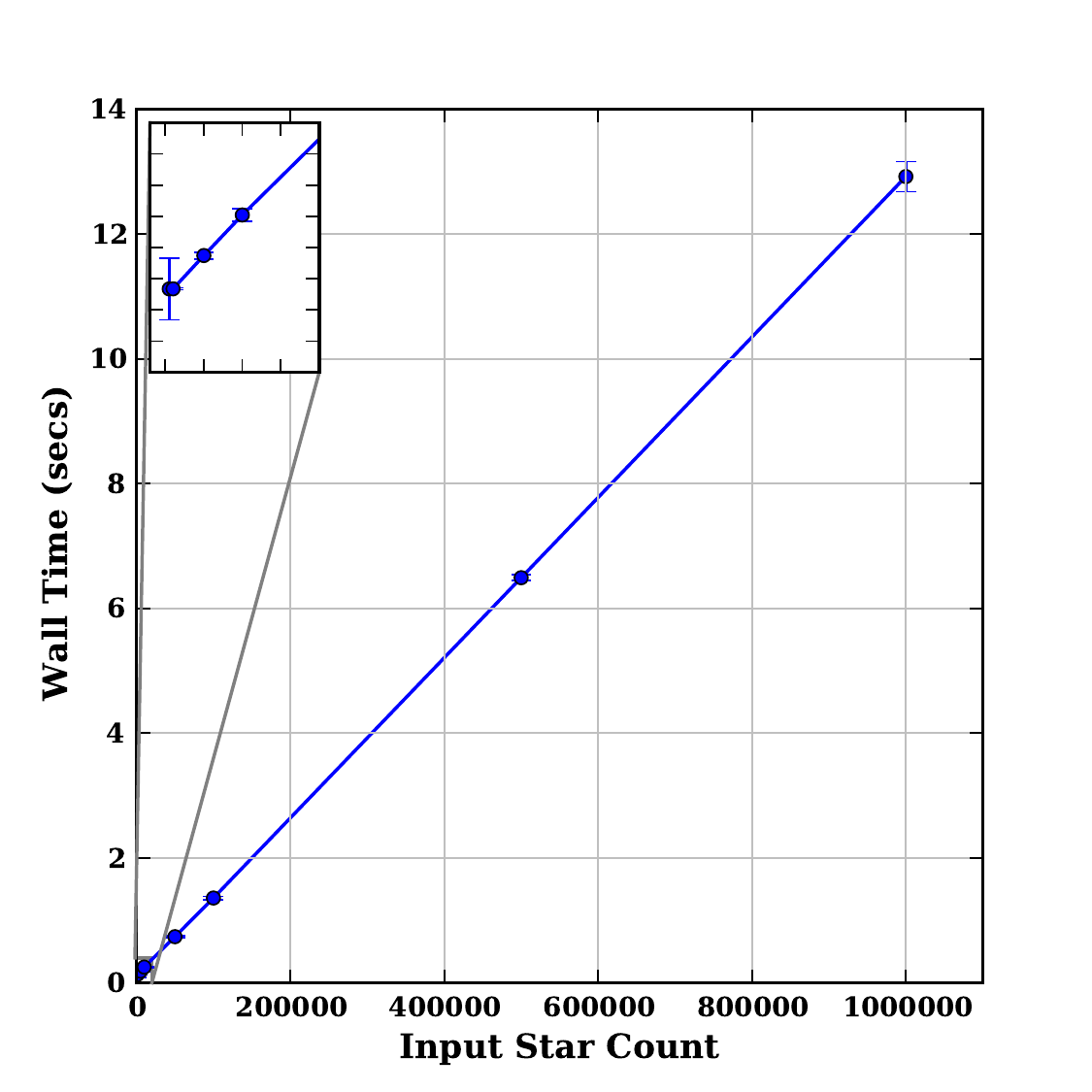}
		\caption{Scalability of our PIX2SKY routine with the number of input elements, for a fixed number of processes. The number of input coordinates was varied from 500 to 1 million. The benchmark was carried out on an AMD quad core machine with 4 parallel processes. The data points are the average of 50 iterations, with error bars corresponding to $1 \sigma$. The inset box is a close-up of the area near the origin. The Process/Queue approach was used with guided scheduler.}
		\label{fig:fig7}
\end{figure*}

Data parallelism using Python multiprocessing scales nicely with the number of processor cores, as shown in Section \ref{examples}. Our coordinate transformation code also scales nicely with the number of input data elements. As depicted in Figure \ref{fig:fig7}, the PIX2SKY routine scales linearly with the number of input data elements. It was benchmarked on an AMD quad core machine for 4 processes, using the Process/Queue multiprocessing method and varying the input data elements from 500 to 1 million. Similar scalability is also achieved for the SKY2PIX routine.

As Python is a platform-independent interpreted language, the parallel code implementation is portable, i.e. it can be implemented on any of the Python-supported OS. In two of our examples, we have used IRAF tasks which can only be executed on Linux or Unix-like OS (e.g. Mac OS X), as IRAF is only available for these platforms. But astronomical data analysis tasks not requiring IRAF can be run on other supported platforms. 

\section{Conclusions}
\label{conclude}
We have shown that moving to parallel astronomical data processing is not as daunting as most astronomers perceive. The open source Python programming language provides the necessary tools to implement parallel code on multicore machines. We used three different hardware configurations, three different parallelizing schemes or approaches, two different load balancing routines, and three different applications of varied complexity to demonstrate the ease of implementation and benefits of parallelizing data processing tasks. Although the emphasis was on the Python \textit{multiprocessing} module, results from the Parallel Python module were also presented. The Process/Queue approach performed better as a parallelizing scheme than both the Pool/Map and Parallel Python approaches. Parallel performance can be optimized by carefully load balancing the workload. Where there is no possibility of re-writing the code for parallel processing because of complexity or any other factor, the existing serial code can still be used to parallelize the problem (as shown in the completeness test and sub-sampled deconvolution tasks). While these are not the only or the most optimized methods to parallelize the code, the computational time savings are still very significant, even with these straightforward approaches. The cross-platform nature of Python makes the code portable on multiple computer platforms.

\section{Acknowledgements}
We gratefully acknowledge the support of Science Foundation Ireland (under award number 08/RFP/PHY1236). Based on observations made with the NASA/ESA Hubble Space Telescope in programs GO-8118 and GO-5366, obtained from the data archive at the Space Telescope Institute. STScI is operated by the association of Universities for Research in Astronomy, Inc. under the NASA contract NAS 5-26555. STSDAS, TABLES, PyRAF and PyFITS are products of the Space Telescope Science Institute, which is operated by AURA for NASA. We also thank the anonymous referees for comments and suggestions, that greatly improved the clarity of the article.

\bibliographystyle{model2-names}
\bibliography{multicore}







\end{document}